# Rolling motion of a rigid sphere on a structured rubber substrate aided by a random noise and an external bias

Partho S. Goohpattader, Srinivas Mettu and Manoj K. Chaudhury*

*Department of Chemical Engineering, Lehigh University, Bethlehem, PA 18015, USA*

**Abstract**. We study the rolling motion of a small solid sphere on a surface patterned rubber substrate in an external field with and without a noise. In the absence of the noise, the ball does not move below a threshold force above which it accelerates sub-linearly with the applied field. In the presence of a random noise, the ball exhibits a stochastic rolling with a net drift, the velocity of which increases non-linearly with the strength of the noise ($K$). From the evolution of the displacement distribution, it is evident that the rolling is controlled by a Coulomb like friction at a very low velocity, a super-linear friction at an intermediate velocity and a linear kinematic friction at a large velocity. This transition from a non-linear to a linear friction control of motion can be discerned from another experiment in which the ball is subjected to a small amplitude periodic asymmetric vibration in conjunction with a random noise. Here, as opposed to that of a fixed external force, the rolling velocity decreases with the increase of the strength of the noise suggesting a progressive fluidization of the interface. Furthermore, the ball exhibits a flow reversal with the increase of the amplitude of the asymmetric vibration, which is indicative of a profile of friction that first descends and then ascends with velocity. An approximate non-linear friction model with the non-linearity decreasing with $K$ is somewhat consistent with the evolution of the displacement fluctuation that also explains the sigmoidal variation of the drift velocity with the strength of the noise. This research sets the stage for studying friction in a new way, in which it is submitted to a noise and then its dynamic response is studied using the tools of statistical mechanics. Although more works would be needed for a fuller realization of the above stated goal, this approach has the potential to complement direct measurement of friction over several decades of velocities and other state variables.

PACS number:

*E-mail: mkc4@lehigh.edu

# I. INTRODUCTION

In a conventional Brownian motion, friction and the thermal noise are coupled to each other. On the other hand, delivering the noise externally by decoupling it from friction allows one to study the nature of the friction itself and to investigate how it affects various dynamics phenomena of interests to adhesion, tribology, wetting and granular flow. We have recently been interested in studying [1-3] the role of non-linear friction in the stochastic motion of a small solid object or a liquid drop on a solid support in the presence of an external noise and a bias. In these cases, the drift velocity increases somewhat sub-linearly, but the diffusivity increases super-linearly with the strength of the noise. The displacement fluctuation exhibits a non-Gaussian behavior at a short time scale, but a Gaussian behavior at a longer time scale. These results can be fairly understood within the framework of a Langevin equation with a Coulombic friction in conjunction with a viscous dissipation. Being inspired by these previous studies, we wished to explore how rolling friction might be affected by a stochastic noise. Additional inspirations for these studies were derived from the fact that non-Gaussian displacement statistics is also observed [4] with a colloidal particle undergoing a Brownian motion in contact with a substrate.

Specifically, we studied the dynamics of a small sphere rolling on a fibrillated rubber [5] surface with the latter being vibrated with a random mechanical noise. A fibrillar surface mimics the features of well decorated asperities with which to study the physics of pinning-depinning transition [6] and bioinspired adhesion [5]. The analysis of the stochastic rolling data required us to conjecture a complex non-linear model of friction with the non-linearity decreasing with the strength of the noise. Additional experiments were designed to interrogate this complexity by subjecting the ball to a deterministic asymmetric vibration and a stochastic noise.

The dynamics of the motion of a line that is pinned randomly by defects [7] is supercritical in the sense that no motion is observed when the applied force is less than a threshold value, above which the velocity ($V$) usually grows non-linearly with the applied field. The extended relationship between the force and velocity can, however, be quite complex. For example, in the peeling of a soft rubbery adhesive from a substrate [8] it is known that the force first increases with velocity and after reaching a peak value, it decreases only to rise again at even higher velocities. The friction of a soft rubber [9-12] on a solid substrate also exhibits a complex velocity-force relationship [13-14] that is similar to that of the peeling of a viscoelastic adhesive from a surface.

Obtaining adequate expressions for velocity dependent friction or crack propagation is not only important for the macro-scale descriptions of these phenomena, they are also critical to the development of appropriate molecular and/or mesoscopic level models of adhesion and friction. During the course of this work, we also explored whether it is possible to obtain an insight into how friction or adhesion depends over a wide range of velocity by subjecting a system to random forces of various magnitudes, and examining its stochastic displacements. The picture of friction that emerged from such a study could serve as a guideline for future experiments with which to explore the dependence of rolling friction on wide range of state variables.

## II. THEORETICAL BACKGROUND

The stochastic motion of an object on a surface in the presence of a Coulombic friction [15] exhibits certain unique characteristics that are different from the dynamics originating from a linear kinematic friction. For example, while the steady state velocity distribution is Gaussian with a linear kinematic friction, it can be exponential or stretched Gaussian with a Coulombic or a non-linear friction. Furthermore, the self-diffusivity with a Coulombic friction depends more strongly on the strength of the noise than is the case with the linear kinematic friction [1]. The problem of Coulombic friction in a stochastic setting was first tackled by Caughey's group [16, 17] at Caltech about fifty years ago, within the framework of a Fokker-Planck equation. Several other studies followed the lead of Caughey [18-20] in the context of the random motion of sliding buildings in response to earthquake. More recently, the problem of stochastic motion involving Coulombic friction has been brought to our attention by de Gennes [21] as well as by Hayakawa et al [22] that received rigorous treatments of path integral [18, 23] and Fokker-Planck [24, 25] formalisms in recent years. The most recent paper by Menzel and Goldenfeld [25] focused on the displacement statistics associated with the random motion governed by Coulombic friction using a Fokker-Planck equation, which was previously addressed using a pulse train excitation approach [26] or a numerical integration of the Langevin equation [1-3]. When both a Coulombic and a viscous friction are at work, Menzel and Goldenfeld [25] showed that the displacement statistics at different time scales are not self similar – it is exponential at short time scale and Gaussian at a longer time scale, which is consistent with experimental observations [1-4]. In spite of the non-Gaussian fluctuation, the variance of the distribution grows linearly [1, 2] at the large time limit. Similar observations were also made by Wang et al [4] in an unusual Brownian motion of a colloidal particle in contact with a microtubule. When a bias is imposed [1, 2], the object drifts with a velocity that increases sub-linearly with the strength of the noise, but linearly with the applied bias. This linear growth of displacement variance with time can be described within the frameworks of a Langevin [1-3] and/or a Fokker-Planck [25] equation.

In order to focus our discussion, let us consider a modified Langevin equation [1-3, 16, 21-27] in which the object experiences a Coulombic as well as a linear kinematic friction:

$$\frac{dV}{dt} + \frac{V}{\tau_L} + \sigma(V)\Delta = \bar{\gamma} + \gamma(t), \qquad (1)$$

Here, $V$ is the velocity of the particle, $\bar{\gamma}$ is the external force divided by the mass of the object, $\tau_L$ is the Langevin relaxation time, and $\gamma(t)$ is the time dependent acceleration of the white noise, the power (or the noise strength) associated with which is $K$. $\Delta$ is the magnitude of the dry friction expressed in terms of the static friction force divided by the mass of the object. If $\Delta$ is smaller than $(\bar{\gamma} + \gamma(t))$, the object moves. On the other hand, if $(\bar{\gamma} + \gamma(t)) < \Delta$, the object may remain stuck to the surface. It will set into motion again if another acceleration pulse $\gamma(t)$ with sufficient energy rescues it from this stuck state. As the non-linear dry (or Coulombic) friction exhibits a jump discontinuity at $V=0$, it is convenient to multiply $\Delta$ with a signum function $\sigma(V)$ which is positive when $V > 0$ and negative when $V < 0$ with $\sigma(0)=0$. Within the above formalism, there is no operational difference between dry friction (solid on solid), wetting hysteresis (liquid drop on solid), or adhesion hysteresis [28] as it appears in rolling motion.

Caughey and Dienes [16] considered Eq. (1) (without the bias and the kinematic friction terms), and its corresponding Fokker Planck equation in order to obtain an expression for the transition probability density in the velocity space. Their results showed that normal diffusive like motion prevails even when the dynamics is governed by the non-linear friction but the diffusivity ($\sim K^3/\Delta^4$) varies more strongly with the power of the noise than the case with a linear kinematic friction ($D \sim K$). Another important finding of Caughey et al [16] is that the transition probability density at the stationary state is exponential with velocity.

Although our recent interests to study the role of non-linear friction in stochastic motion stem from its relevance to the problems of soft matter physics, the early interests in this subject arose from its importance in studying the sliding of the building foundations in response to earthquake. In this arena, following the lead of Caughey and Dienes [16], Ahmadi [19] and Crandall et al [20] presented some approximate, but useful results. Below, we briefly review and extend certain predictions of Caughey and Dienes [16] as well as those of de Gennes [21] and Hayakawa et al [22] which would be important in interpreting the results of the experiments performed by us.

The non-linear nature of Eq. (1) makes it cumbersome to treat it analytically. As far as average values are concerned, one way to tackle the problem is to consider a classical linear version [16, 17] of this equation and estimate the equivalent of the Langevin relaxation time. Following Caughey [16, 17] and Crandall et al [20], we express Eq. (1) (without the bias) in the form shown in Eq. (2) with the addition of a remainder term $\varphi$ as in Eq. (3).

$$\frac{dV}{dt}+\frac{V}{\tau_L^*}=\gamma(t), \tag{2}$$

$$\varphi=\frac{V}{\tau_L^*}-\frac{V}{\tau_L}-\sigma(V)\Delta, \tag{3}$$

The criterion for equivalent linearization is to minimize the average value of $\varphi^2$ with respect to $\tau_L^*$, which leads to the following equation:

$$\frac{1}{\tau_L^*} = \frac{1}{\tau_L} + \frac{\Delta \langle \sigma(V)V \rangle}{\langle V^2 \rangle}, \tag{4}$$

Calculation of the averages shown in Eq. (4) requires an expression for the stationary probability density of velocity. This can be obtained by setting the diffusive flux in the velocity space to zero:

$$\frac{K}{2}\frac{\partial V}{\partial V} + \Delta \frac{|V|}{V}P + \frac{VP}{\tau_L} = 0, \tag{5}$$

The stationary velocity distribution (P(V)) is

$$P(V) = P_o \exp\left(-\frac{V^2}{K\tau_L} - \frac{2|V|\Delta}{K}\right), \tag{6}$$

The averages in Eq. (4) are now carried out with the velocity distribution function given in Eq. (6). The analysis can be simplified if the exponential term (due to Coulombic friction) of Eq. (6) dominates over the kinematic term, which is often the case. One thus obtains:

$$\frac{1}{\tau_L^*} = \frac{1}{\tau_L} + \frac{\Delta^2}{K}, \tag{7}$$

Eq. (7) defines the equivalent relaxation time in terms of the Coulombic and a linear kinematic friction. When a small external bias is imposed, an expression for the drift velocity [1, 2, 21, 27] can be obtained from the linear response theory, i.e. $V_{drift} = \bar{\gamma}\tau_L^*$. We thus have,

$$V_d = \frac{\bar{\gamma}\tau_L}{1 + \Delta^2 \tau_L / K}, \tag{8}$$

In a typical experiment of stochastic rolling or sliding, one can perform two types of measurements. With a random noise and a bias, the ball rocks forward and backward randomly but with a net drift. At a given bias, one can record the motion of the ball over a large distance for a given duration of time and estimate the drift velocities. Alternately, one can record the stochastic motion of the ball with a high speed camera to study the trajectory over certain duration of time. The spatial segments of the trajectories corresponding to a given time segment can then be used to obtain a PDF of the displacement fluctuation. Such a pdf has a given peak and a dispersion of displacements. By plotting the position of the peak as a function of time segments, a drift velocity can be estimated. Furthermore, from the slope of the variance of the displacement versus time, one can obtain the diffusivity. When only a linear friction operates, the drift velocity should simply be a product of the bias $(\bar{\gamma})$ and the Langevin relaxation time $(\tau_L)$. On the other hand, when only the dry friction operates, the drift velocity is given by $\bar{\gamma} K / \Delta^2$. With the presence of both the kinematic and a dry friction, the drift velocity starts [1, 2, 27] from a very low value and progressively saturates to $\bar{\gamma}\tau_L$ sub-linearly. These predictions are consistent with our previously reported sliding experiments [1, 2], but not, exactly, with a steel ball rolling on a fibrillar PDMS substrate. Here the drift velocity increases in a sigmoidal fashion with the strength of the noise. Understanding this discrepancy is the central objective of this paper.

At this point we should mention that a non-linear evolution of the drift velocity with K can also be observed with a non-linear friction of the type: ($\sim |V|^n$). Here, the Langevin equation is:

$$\frac{dV}{dt} + \frac{B|V|^n}{M}\sigma(V) = \gamma(t), \tag{9}$$

The stationary probability distribution function [1, 29] for the velocity is given by the following equation:

$$P(V) = P_o^{"} \exp\left(-\frac{2B|V|^{1+n}}{M(1+n)K}\right), \tag{10}$$

Eq. (10) suggests that the velocity pdf is exponential if the friction is Coulombic (i.e. $n=0$). It is Gaussian for a linear kinematic friction with $n=1$. A super or a sub Gaussian pdf results for $n>1$ and $n<1$ respectively.

Using the method of the equivalent linearization, it is easy to show that the characteristic relaxation time ($\tau*$) scales as $K^{\frac{1-n}{1+n}}$, whereas $\langle V^2 \rangle$ scales as $K^{\frac{2}{1+n}}$ yielding drift velocity and diffusivity [30] growing with K as $\bar{\gamma} K^{\frac{1-n}{1+n}}$ and $K^{\frac{3-n}{1+n}}$ respectively. The exact reproduction of the values of diffusivity that would match the experimental results [1] is not, however, an easy task although the experimentally observed exponent of *K* is satisfactorily explained. The main difficulty lies in the lack of adequate knowledge of the correlation of the velocity and displacement fluctuations, the origin of which remains somewhat elusive in systems governed by non-linear friction with the possibility of trapped states. With the aid of random trajectories from a given solution of a non-linear Langevin equation, and subsequently destroying the correlation, it is possible to show that different evolution paths of the variance leads to different diffusivities. The values of the drift velocities, however, remain rather robust. The best we can do at present is to make qualitative assessments of the nature of friction that contributes to the shape of a displacement pdf and then use this insight to predict drift velocities. By focusing on the small deviation behavior of displacement fluctuations we gain insights into the nature of frictional dynamics in the small velocity region, whereas the large deviation behavior of displacement fluctuation provide information of such a dynamics contributed by the large velocities underlying an atypical Brownian motion.

## III. EXPERIMENTAL SECTION

The instrument employed for this study is an inertial tribometer, which was used previously first by Baumberger et al [31] and later by others [1, 2, 32]. When the substrate is slightly inclined ($< 3^o$) from the horizontal, no motion of the ball is observed as the force needed to break the adhesive junction is greater than that can be provided by gravity. This is similar to the Coulombic friction preventing the sliding of a solid object, or the wetting hysteresis preventing the rolling of a liquid drop, on a surface. The ball, however, rolls at an inclination less than the threshold value if it is subjected to an external vibration. When the vibration pulses are random, the motion of the ball resembles that of a drifted rotational Brownian motion. We measured both the drift velocity as well as studied the displacement fluctuation of the ball submitted to a random Gaussian noise. Typical experiment was to place a small steel ball (4 mm diameter) on a slightly inclined ($1^o$) fibrillated PDMS film and subject the latter to a random vibration (Fig. 1). Although most of our experiments were conducted with a Gaussian random noise, some of the experiments were performed with an asymmetric vibration with or without the noise. The substrate was attached to an aluminum platform connected to the stem of a mechanical oscillator (Pasco Scientific,

Model SF-9324). Gaussian white noise was generated with a waveform generator (Agilent, model 33120A) and fed to the oscillator via a power amplifier (Sherwood, Model No: RX-4105). By controlling the amplification of the power amplifier, noises of different powers were generated while keeping the pulse width constant at ~ 40 *μs*. The acceleration of the supporting aluminum plate was estimated with a calibrated accelerometer (PCB Peizotronics, Model No: 353B17) driven by a Signal Conditioner (PCB Peizotronics, Model No: 482) and connected to an oscilloscope (Tektronix, Model No. TDS 3012B). The PDFs of these accelerations [1] are Gaussian and their power spectra are flat up to a total bandwidth of 10 *kHz*. The entire setup was placed on a vibration isolation table (Micro-g, TMC) to eliminate the effect of ground vibration. The motion of the ball was recorded with a high speed camera (Redlake, MotionPro, Model 2000) operating at 1000 frames/sec. Motion analysis software MIDAS was used to track the dynamics of the steel ball.

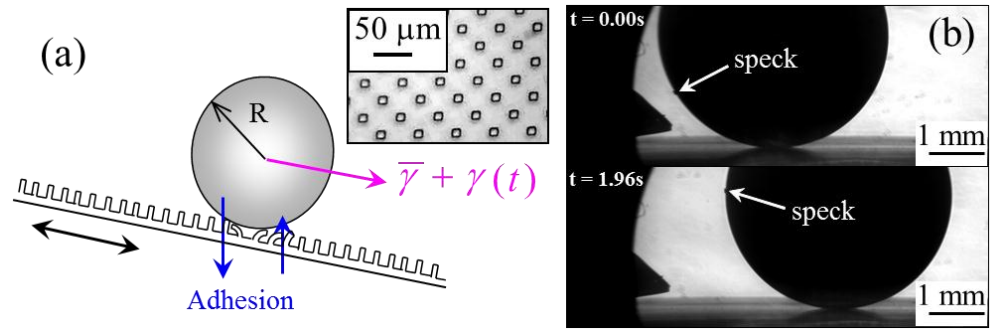

FIG. 1. (Color online) (a) A steel ball of diameter 4 mm rolls on a fibrillated rubber surface at an inclination of 1°. When the moment of the gravitational plus the stochastic force about the point of contact is greater than the torque of adhesion, the ball rolls on the surface. (b) Figure shows a speck of dust moving along the perimeter of the ball by 1.23 mm, which is almost same as the lateral displacement of the ball indicating that the ball undergoes a net rolling instead of sliding at the macroscopic scale. This experiment was performed at noise strength of 0.06 m$^2$/s$^3$. Inset of Fig. (a) shows microscopic image (top view) of the fibrillated PDMS surface.

Micro-fibrillated PDMS (Dow Corning Sylgard 184) surfaces were used as a substrate for the rolling experiment. The preparation of such surfaces is reported in detail elsewhere [4]. Briefly, the oligomeric component of the Sylgard 184 kit was thoroughly mixed with the crosslinker in a 10:1 ratio by weight followed by degassing it in vacuum for 2hrs. The degassed mixture was then cast onto lithographically etched silicon master. These master wafers were silanized for easy removal of cured fibrillated PDMS sample. The cast PDMS was then cured at 80°C for 2hrs. The crosslinked PDMS was cooled in dry ice (-78.5°C) for an hour followed by its removal from silicon master wafers manually. The PDMS surface thus prepared has square fibrils of 10$\mu m$ size with a center to center distance of the adjacent fibrils of 50 $\mu m$. The height of the fibrils was 25$\mu m$.

The steel ball used in our experiment was a bearing quality aircraft grade E52100 steel obtained from Mcmaster corporation (http://www.mcmaster.com/#chrome-steel-balls/=cph9ai ). The diameter of the ball was 4 mm with a tolerance of $\pm 2.5 \mu m$. The balls were cleaned by sonicating it in acetone and then drying in nitrogen. The root mean square roughness of the surface of the steel ball was ~ 35 nm as obtained from atomic force microscopy (Veeco nanoscopeV, Digital Instruments, Metrology Group) over a scanning area of 400 $\mu m^2$.

The rolling experiments were carried out at 19 different strengths (or powers) of the noise ranging from 0.02 $m^2/s^3$ to 2.7 $m^2/s^3$ at a bias of 0.04 mN. The angle of inclination was controlled with a precise goniometer (CVI Melles Griot, Model No: 07 GON 006).

## IV. RESULTS

### A. Rolling of the steel ball on the fibrillated rubber without noise

We first describe the results of the rolling of a steel ball on the fibrillated rubber without the noise. The ball starts rolling on the rubber at an angle of about 3°. A video microscopic image of the motion of the ball shows that it accelerates as it rolls down. The fact that the data can be fitted with a simple equation of accelerating motion of the type $S = V_i t + 0.5\ at^2$ ($V_i$ being the initial velocity and $a$ is the falling acceleration) suggests that there is virtually no kinematic friction acting on the ball. Only resistance here is a Coulombic type dry friction. If this is the case, the acceleration should simply be $mg\sin\theta - \Delta$. Experiments carried out at different angle of inclination show that beyond a threshold angle ($\theta_c$), the acceleration increases (Fig. 2) with the angle of inclination as, $a \sim (\sin\theta - \sin\theta_c)^{2/3}$. This sub linear growth of acceleration with the angle of inclination suggests that the dry friction resistance ($\Delta$) increases with the applied force as well. In the past, it has been proposed [7, 33] that the dynamics of the motion of a line that is pinned randomly by defects exhibits a supercritical behavior in the sense that no motion is observed when the applied force ($F$) is less than a threshold value ($F_c$), above which the velocity ($V$) grows as $V \sim (F-F_c)^\theta$. $\theta$ is the velocity exponent, the value of which lies in the range of 0.6 to 0.8. We are not aware of any analysis suggesting the strengthening of dry friction with force. The observed effect may suggest a force induced broadening of the contact line zone where random pinning of the interface occurs. The behavior of the fibrillated surface with a noise reveals that the surface bristles get considerably fluidized ensuring the emergence of a kinetic friction.

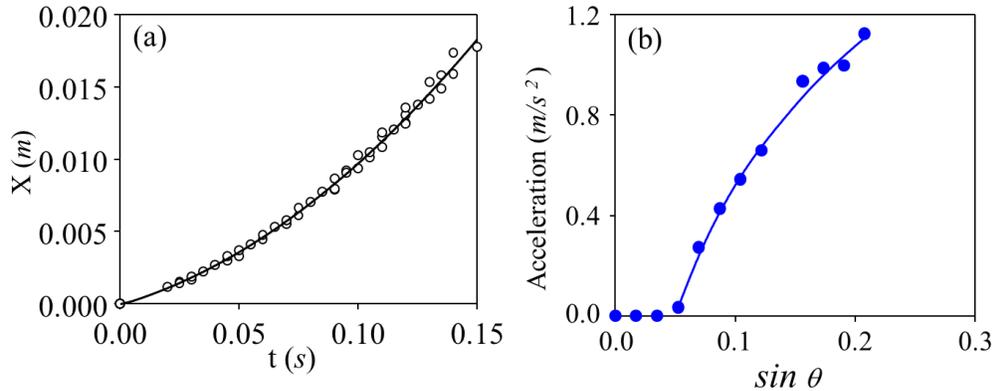

FIG. 2. (Color online) (a) Figure 2(a) shows a parabolic growth of the distance travelled by a ball on an inclined (10°) surface with time. The falling accelerations are summarized in Fig. 2(b).

### B. Stochastic motion of the steel ball

When the substrate is vibrated with a Gaussian noise, the steel ball undergoes backward and forward stochastic rolling motions with a net downward drift. In order to ensure that the ball indeed rolls on the surface, we examined few video clips carefully with a small speck of dust on the surface of the ball (Fig. 1(b)). At a low intensity of noise (0.06 $m^2/s^3$), the speck moved on the surface by the same distance as the ball's lateral displacement. At the high intensity of noise (1.7 $m^2/s^3$), the movement of the speck on the surface of the ball was 3% to 10% lower than its lateral displacement. These measurements ensured that ball undergoes a net rolling motion on the substrate on the average, even though some sliding may be occurring at a stochastic time scale. The displacement of the ball is linear with time, suggesting that it is controlled by viscous like friction. Prandtl [34, 35] pointed out almost 100 years ago that the frictional response of a system changes from Coulombic to kinematic in the presence of a thermal noise (M. Muser, personal communication, 2010). So, we must clarify what we mean by the role of the viscous like friction. These points can be further elucidated by examining the details of how the drift velocity varies with the strength of the noise and how the displacement fluctuation grows with time. However, before a quantitative discussion of the nature of the drift observed with the random noise, it is important to understand the nature of the noise that is used in these studies.

### *1. Characteristics of the noise*

The approximate Eq. (8) is derived on the condition that the noise is strictly white and Gaussian. This is not exactly the case for the type of noise that we generate experimentally. In our case, the Gaussian white noise is generated with a waveform generator that feeds pulses of random heights, but of a finite width (40 $\mu s$) to an oscillator. The output of these pulses is used to vibrate the stage on which the rolling experiment is performed. Since a mechanical oscillator has a tendency to spring back after each excitation, the autocorrelation of the output noise exhibits a negative peak (Fig. 3(b)), which is also consistent with the ACF of the noise generated numerically using the properties of the oscillator (Fig. 3(c)). The noise pulses, however, are Gaussian with a probability density of $P = P_o \exp\left[-0.5(\gamma/\sigma_\gamma)^2\right]$, as evidenced from the slope (~2) of the plot of $\ln(-\ln(P/P_0))$ versus $\ln|\gamma/\sigma_\gamma|$ (Fig. 4). Because the noise is somewhat correlated, Eq. (8), which is derived on the basis of the classical Fokker Planck equation, needs to be corrected. This correction is carried out as follows. By numerically integrating the Langevin equation (Eq. (1)) with the omission of the kinematic friction term and using the sequence of the noise pulses obtained directly from the accelerometer, several trajectories are generated. The strength of the noise as used in these experiments is nominally defined as the product of the mean square acceleration and the pulse width ($\tau_c$), i.e. $K = \langle \gamma(t)^2 \rangle \tau_c$. However, the value of this $K$ is re-normalized in order to use it in Eq. (8). For a given set of $\Delta$ and $K$, 100 trajectories, each lasting for 6 seconds, were used to estimate the drift velocity. Although this drift velocity varies (Fig. 5) linearly as $\bar{\gamma} K / \Delta^2$, its slope is found to be 0.03. Thus, Eq. (8) is modified as:

$$V_d = \frac{\bar{\gamma} \tau_L}{1 + 33 \Delta^2 \tau_L / K}, \tag{11}$$

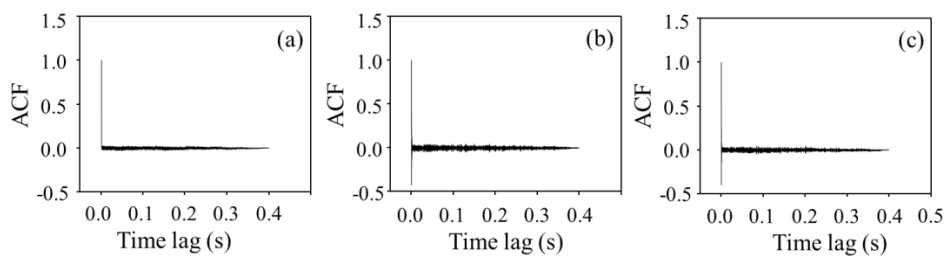

FIG. 3. The autocorrelation of the noise file (3(a)) as generated from the computer and that (3(b)) obtained from the output of the oscillator as measured with an accelerometer. The Gaussian noise as generated from the waveform editor, $\gamma(t)$, was used to solve the Langevin equation of the oscillator: $\ddot{x} = -\dot{x}/\tau - \omega_o^2 x + \gamma(t)$. Here, x is the displacement of the oscillator, $\tau$ (250 $\mu s$) is its relaxation time and $\omega_o$ (~1.5x10$^4$ s$^{-1}$) is its fundamental frequency of vibration. The autocorrelation of the simulated noise of the acceleration is shown in Fig. 3(c).

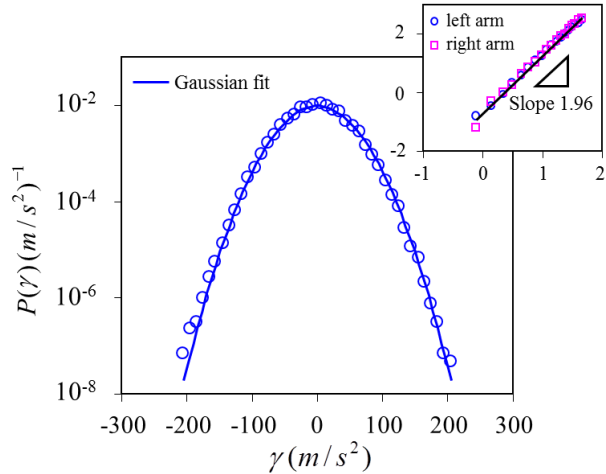

FIG. 4. (Color online) Probability distribution function of the noise obtained from accelerometer at a given value of $K$ (0.06 m$^2$/s$^3$). The pdf is also fitted with a Gaussian function as indicated by the solid line. The inset shows the plot of $\ln(-\ln(P/P_o))$ versus $\ln|\gamma/\sigma_\gamma|$, the slope of which is ~2.

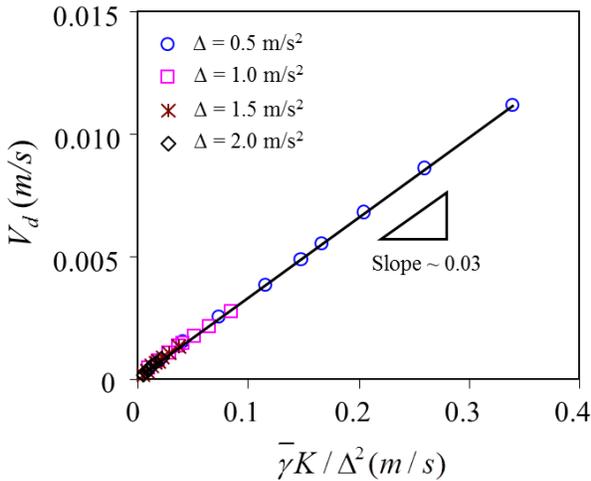

FIG. 5. (Color online) The drift velocity is calculated using Eq. (1) without the kinematic term using the noise output file of an accelerometer attached to an oscillator. Various values of $\Delta$ are used. A master plot is obtained by plotting all the drift velocity data against $\bar{\gamma} K / \Delta^2$.

## *2. Drift velocity and the strength of noise*

The steel ball rolled on a straight path without exhibiting any significant sidewise drift. Using a low magnification camera, the drift velocities were obtained from the displacement of the ball for a given duration of time using several tracks for each noise strength. The stochastic displacement of the ball was also examined in detail with a high speed camera. At a low power (0.06 m$^2$/s$^3$), each track lasted for about 6s. This track was divided into different time segments (0.001s to 1s) using all possible starting and ending times. By combining data obtained from all the tracks, the displacement pdfs were constructed for different durations of time. Each pdf exhibits a certain peak and a variance. The drift velocity obtained from the time dependent shift of the mean is same as what is measured with a low resolution camera. In order to examine the behavior of the displacement pdfs at a high power (1.7 m$^2$/s$^3$) a larger number (200) of tracks was used.

The drift velocity of the steel ball increases (Fig. 6) with the strength of the noise ($K$) and tends to saturate at high $K$. This observation is similar to our earlier observations with a noise induced sliding of a small solid object or a small water drop on a surface [1, 2]. However, unlike the previous observations, the $V_d$-$K$ curve here has a knee at low $K$ thus exhibiting a slight sigmoidal behavior. All the data can be fitted well with an equation of the type: $V_d = \bar{\gamma}\tau_L \tanh(K/K_1)^{1.4}$ using a value of $K_1$ as 0.7 m²/s³. The unique $V_d$ vs $K$ relationship clearly suggests that a non-linear friction is operative underlying the rolling motion. There is no definite time scale to the problem except at very high $K$, where the saturation of the velocity implies a Langevin time scale of ~ 0.1s. A fit (Eq. (11)) of the $V_d$-$K$ curve by keeping with the fact that that the velocity goes as $0.03\bar{\gamma} K/\Delta^2$ at very low values of $K$, and it approaches $\bar{\gamma}\tau_L$ at high values of $K$, exhibit the sublinear evolution of drift velocity as shown in Fig. 6. Although this fit does not reproduce the sigmoidal behavior seen experimentally, it is consistent with the fact that the drift velocity is controlled by a non-linear friction at low noise strength but by a linear viscous friction at high noise strength. In order to glean further insights into this complex friction dynamics, let us now examine the evolution of the displacement fluctuation of the steel ball obtained at a low and a high strength of the noise.

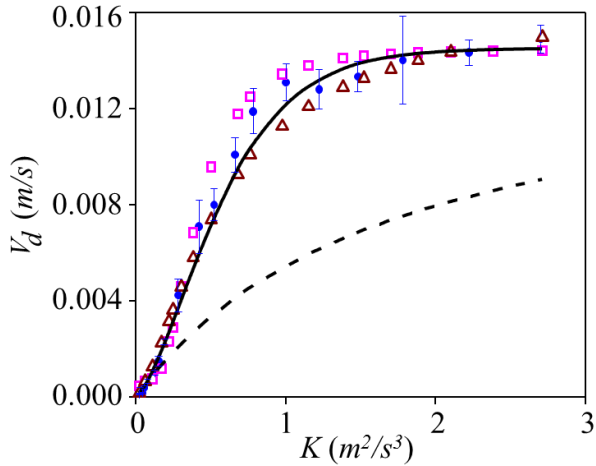

FIG. 6. (Color online) Drift velocity increases with the power of the noise. The filled blue circles are the experimental data. The dashed line represents the velocity obtained using Eq. (11). In order to construct this plot particular values of $\Delta$ and $\tau_L$ had to be used. The value of $\Delta$ (0.8 m/s²) was obtained by fitting the drift velocity with $\bar{\gamma}K/\Delta^2$ at the very low values of $K$, $\tau_L$ (0.1 s) was approximated from the saturated value of the drift velocity. Solid line represents the velocity obtained using equation $V_d = \bar{\gamma}\tau_L \tanh(K/K_1)^{1.4}$. The open squares and triangles represent the data obtained using the three state and two state models of friction (see below).

## C. The nature of non-linear rolling friction as gleaned from the displacement fluctuations

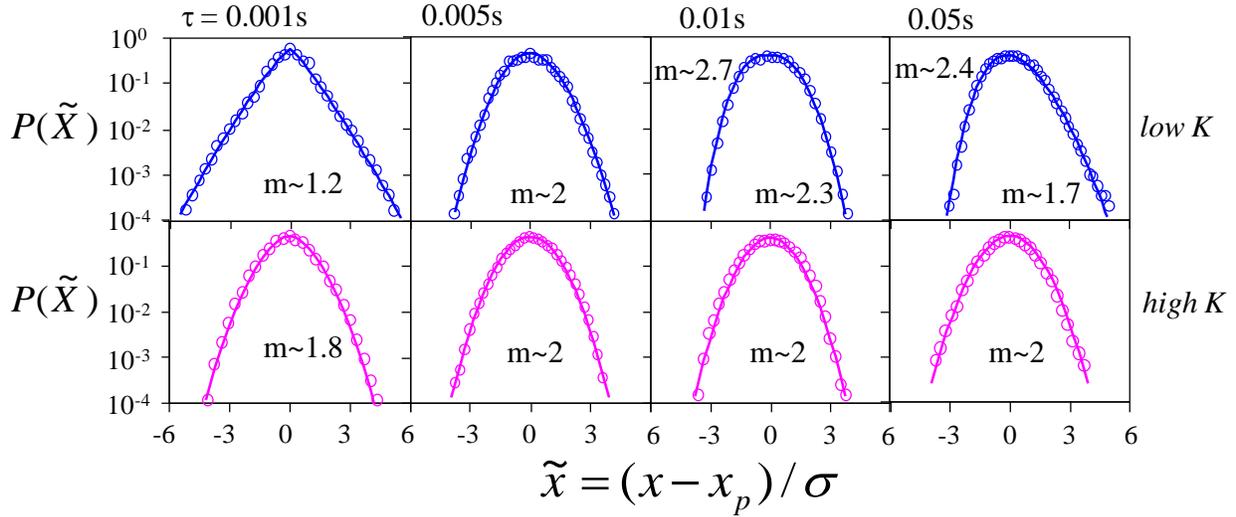

FIG. 7. (Color online) Summary of the fluctuations of the displacements of a steel ball rolling on a fibrillated PDMS at a bias of 0.04 mN corresponding to the time segments of 0.001s, 0.005s, 0.01s, and 0.05s respectively. Low $K$ and high $K$ correspond to 0.06 m$^2$/s$^3$ (upper panel) and 1.7 m$^2$/s$^3$ (lower panel) respectively. The pdfs are fitted as $P \sim \exp(-c|\tilde{x}|^m)$, with the values of $m$ embedded inside the figures. For a symmetric pdf, only one value of $m$ is given. For an asymmetric pdf, two values of $m$ are given, one for the left and the other for the right side of the pdf.

Figure 7 summarizes the fluctuations of the displacements of the steel ball on the fibrillated PDMS substrate corresponding to a low bias (0.04 *mN*) at two different noise strengths. It should be borne in mind that these displacement pdfs bear the signatures of velocity dependent friction. The displacement pdf for $K$=0.06 m$^2$/s$^3$ at $\tau$=0.001s (Fig. 7) is much sharper than that would be expected of a Gaussian behavior. This supports the picture that a friction resembling dry friction operates near the zero velocity region. The pdf for $\tau$=0.01 does not show the sharpness of the peak that is observed for $\tau$=0.001s. This pdf is, superficially, quite Gaussian thus suggesting that a viscous friction operates at higher velocity. The pdf corresponding to $\tau$=0.05 s also appears to be Gaussian, but it is somewhat asymmetric.

More detailed information regarding the natures of these pdfs can be surmised by considering the velocity distribution as given in Eq. (10). As a consequence of a stretched Gaussian velocity distribution, the displacement fluctuation at short time limit should also follow a function of the type, $P = P_O \exp\left\{-c\left[(x-x_p)/\sigma\right]^m\right\}$, where $\sigma$ is the width of the pdf and $x_p$ is the displacement corresponding to the peak of such a distribution. In the case of a power law type friction, it is, however, not easy to define the stationary state unlike the case with that of a linear friction; which is also dependent on the strength of the noise. Thus, although we expect that the exponent of the above stretched Gaussian distribution of the displacement pdf to be $n+1$, we need to support this conjecture by a numerical integration of the Langevin equation. Numerical integration of Eq. (15) was carried out using a generalized integration method for stochastic differential equations as outlined by Gillepsie [36]. Stochastic acceleration of the vibrating plate as measured using an accelerometer were used as the input, $\gamma(t)$, in the same sequence as they were generated experimentally to ensure that the noise correlation is identical in experiment and simulation. While the simulated drift velocity as well as the variance of the displacement did not depend on the integration time step (20$\mu s$–80$\mu s$), all the simulations were carried out with an integration step of 20$\mu s$. Eq. (13a) with a bias of 0.04mN was integrated numerically using an expression of friction as $f(V) = A|V|^n$ with the value of $n$ as 0.2, 0.5 and 1.0 respectively for a value of K=0.06m$^2$/s$^3$. The value of $A$ was so chosen that the average velocity obtained from simulations consisting of 100 tracks, each being 6s long, matches the experimental value (0.67 mm/s). The pdfs of displacements constructed from such stochastic trajectories are shown in Fig. 8 for a time scale of 0.01s. These data were further analyzed by plotting $\ln(-\ln(P/P_o))$ vs $\ln|(x-x_p)/\sigma|$. The slopes of these plots are about 1.2, 1.5 and 2.0, for $n$=0.2, 0.5 and 1 respectively suggesting that the displacement pdfs are stretched Gaussian in general with an exponent ($m$), the value of which is greater than the velocity exponent ($n$) by unity. Figure 8 shows that these stretched Gaussian functions are consistent with the experimental data.

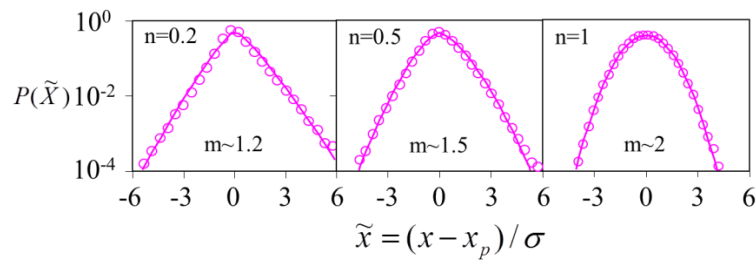

FIG. 8. (Color online) Simulated pdf of displacement for a time segment of 0.01s as obtained from the numerical integration of the Langevin equation using a non-linear friction law: $f(V) \sim |V|^n$. The pdfs are fitted as $P \sim \exp(-c|\tilde{x}|^m)$, with the values of $m$ embedded inside the figures.

In addition to the small deviation behavior of the displacement pdf, which provides insights into the nature of friction at the low velocity range, we can also examine its large deviation behavior in order to gain insights into its large velocity dynamics. This is so, as a large velocity fluctuation of low probability inserts a large displacement to the trajectory that would be evident in the pdfs of a larger timescale. When the data of Fig. 7 are analyzed in the same way as that of Fig. 8, it is found that for small time scale $\tau = 0.001$s, the exponent of the displacement pdf is 1.2. The pdf becomes Gaussian at a time scale of 0.005s, but super-Gaussian at a larger time scale. Although this pdf is somewhat asymmetric, its average exponent (~2.5) is larger than 2 suggesting that the dynamics is governed by a super-linear friction with a velocity exponent of about 1.5. The average exponent, however, tends to the value of 2 expected of the linear kinematic friction at a large value of $\tau$. In our previous studies with sliding motion [1, 2], we observed such an asymmetry as well, with the skewness increasing with the bias and the non-linearity of friction. In the present context, the observed asymmetry may also result from a sublinear friction following a super-linear behavior.

In view of the evidences gathered so far, we arrive at this scenario. A Coulombic type friction operates in the low velocity region followed by a super-linear ($n >1$) friction at a larger velocity range. Finally, the friction becomes linear kinematic at a much larger velocity region. In going from a superlinear to a linear behavior, the friction has to overcome a hump that mimics the adhesive peeling behavior [8] from solid surfaces.

### D. Evidence gathered from the rolling motion with an asymmetric vibration

Another evidence of the complex nature of rolling friction can be gleaned from an experiment, in which the steel ball is submitted to an asymmetric vibration [37-39] of the type:

$$\gamma(t) = A_o \left[ |\sin(2\pi\omega t)| - B \right], \qquad (12)$$

This acceleration has a cusp shape on one side (inset, figure 9(a)), but smoothly varying on the other. The mean value of $\gamma(t)$ is zero. When excited with this waveform, the steel ball moves on the horizontal surface of the fibrillated PDMS (Fig. 9). For any type of motion to occur under a periodic forcing, some kind of non-linearity is required in order to break the symmetry of the applied force. Hence, the observation of the motion of the steel ball by the asymmetric vibration suggests that the friction is non-linear.

#### *1. Flow reversal with asymmetric vibration*

The more interesting observation, however, is the flow reversal that is observed with the increase of the amplitude of the asymmetric vibration at a constant frequency. Let us consider an equation of motion of the type:

$$\frac{dV}{dt} + \sigma(V) f(|V|) = \gamma(t), \qquad (13a)$$

If we integrate Eq. (13a) over one period ($T$), we have the velocity as

$$V(T) = -\int_0^T \sigma(V) f(|V|), \qquad (13b)$$

If the friction is linear in velocity, i.e. $f(|V|) \sim |V|/\tau_L$, it will have the same form as the forcing function that leads to $V(T)=0$. Finite velocity is obtained if $f(|V|)$ is non-linear [37-42]. By solving Eq. (13b) in conjunction with a form of friction that is either constant ((i.e. $f(V) = \Delta$), descends with velocity [i.e. $f(|V|) \sim \exp(-|V|/V_o)$], or sublinear (i.e. $f(|V|) \sim |V|^{0.5}$) it is found that the ball moves from right to left as is found experimentally (Fig. 9). On the other hand, with a friction that strongly increases with velocity (i.e. $f(V) \sim V^{1.5}$), the ball moves from left to right as in Fig. 9. The evolution of the displacement pdfs suggests that friction first decreases and then increases with velocity. With such a pattern of friction, flow reversal can indeed occur as the increasing amplitude of vibration probes sequentially the lower to higher velocity profile of friction. Now increasing the acceleration further, even a larger velocity region is explored where the friction again descends with velocity with which we expect another flow reversal. Such a flow reversal is indeed observed experimentally; however, the motion of the ball in this region is not smooth anymore (Fig. 9).

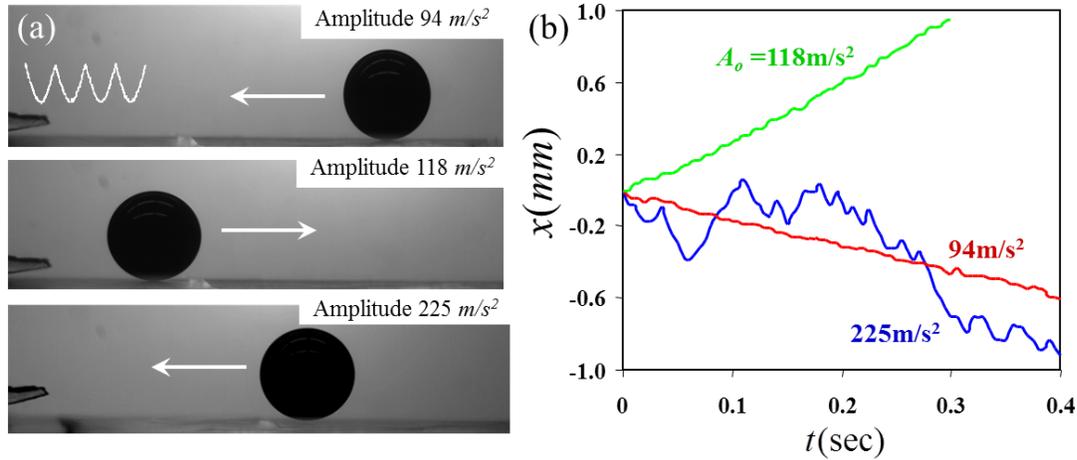

FIG. 9. (Color online) (a) Directions of motion of a steel ball (4 mm) on a horizontal fibrillated PDMS surface at different amplitudes of an asymmetric vibration are shown by the arrows. A typical wave form of the asymmetric vibration of the plate is show in the inset. The frequency of the vibration ($\omega$) is 100 Hz. The directions of motion at different amplitudes are shown by arrows. (b) The displacement of the ball at three different amplitudes of asymmetric vibration are shown as a function of time.

Next we find out what happens when the asymmetric vibration is used in conjunction with a stochastic noise. Here, the noise defines the effective temperature of the system, whereas the asymmetric vibration interrogates it by subjecting it to a rate. What is observed with a low strength noise is that the rolling speed initially increases from that obtained with $K=0$. Beyond a value of $K \sim 0.14 \text{m}^2/\text{s}^3$, rolling speed decreases (figure 10) with the strength of the noise $K$ reaching a very small value of drift velocity ~

0.1mm/s at high value of $K \sim 1.5 m^2/s^3$. This experiment clearly proves the point that the effect of non-linear friction decreases at high $K$, and the system tends towards a nominally fluidized state.

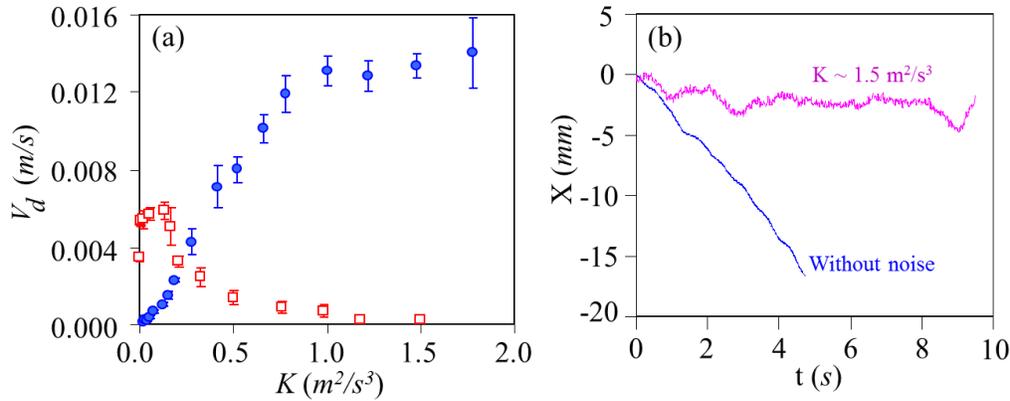

FIG. 10. (Color online) (a) The drift velocity as a function of the strength of the noise for an asymmetric periodic bias (open red square) and a fixed bias (filled blue circle). (b) The trajectories of the ball with and without the noise, but with the asymmetric vibration are shown.

These experiments have interesting similarities to some earlier observations [43, 44] where a granular medium was fluidized with a strong vibration. When the response of the system to an effective temperature ($K$) was interrogated by measuring its friction dynamics, it was observed that a solid like behavior persists at a low state of vibration, but it gets fluidized at high strength of the noise.

### E. A toy model of non-linear friction

The above experiments lead to the following scenario. The friction is non-linear at low velocity, but it becomes linear at high velocity. Secondly, the friction does not vary monotonically with velocity; it exhibits descending and ascending branches as velocity is increased. Thirdly, there is clear evidence of the effect of the noise strength on the overall state [45] of the system, i.e. the system remains in a fluidized state at all velocities when the noise is strong. These are the main findings of this work and any further progress in this research should rest on direct experimentation to obtain the friction force $f(V,K)$ that depends on effective temperature and rate of the system along with a molecular/mesoscopic level understanding of the phenomena. However, a toy model of friction can be constructed that is consistent with the essential features of the displacement pdfs as well as the noise dependent evolution of the drift velocity. In order to illustrate this point, we numerically integrate the Langevin equation (Eq. (15)) of the steel ball with a friction law (Eq. (14)) as follows:

$$f(|V|, K) = f_1(|V|)\exp\left(-\frac{K}{K_1}\right) + \frac{|V|}{\tau_L}, \tag{14a}$$

$$f_1(|V|) = \Delta \exp(-\frac{|V|}{V_o}) + A \exp\left[-\left(\frac{|V|-V_t}{\sigma_V}\right)^{1.5}\right], \tag{14b}$$

$$\frac{7}{5}\frac{dV}{dt} + \tanh(\alpha V) f(|V|, K) = \bar{\gamma} + \gamma(t), \tag{15a}$$

$$\frac{dx}{dt} = V, \tag{15b}$$

In Eq. (14a) and (14b), $f_1(|V|)$ and $|V|/\tau_L$ are the non-linear and linear components of the friction respectively. The fact that the pdf for τ=0.001s at the lower power is sharp, but is nearly Gaussian at the higher power suggests that the non-linear part of friction progressively dies out with increasing $K$, which is captured by the term $\exp(-K/K_1)$. The value of $K_1$ is taken to be 0.7 m²/s³, which is same as that used to fit the drift velocity data using the empirical equation: $V_d = \bar{\gamma}\tau_L \tanh(K/K_1)^{1.4}$. The non-linear friction itself has an exponential term coupled to the dry friction Δ indicating that its effect decreases with velocity, as we have seen in the pdf at the low noise strength. We should point out that an exponential form of the velocity weakening Coulombic friction is within the scope of the current treatment of solid friction [46]. The value of Δ (0.8 m/s²) is close to that obtained from fitting the drift velocity to $0.03\bar{\gamma} K/\Delta^2$ in the very low $K$ limit. The second term of the right hand side of Eq. (14b) is a stretched Gaussian with an exponent of 1.5 that reflects the broadening of the displacement pdf at the intermediate time scale (Fig. 11). The distribution is centered around $V_t$ ~ 0.012 m/s, which is similar to our previous observations [12] of sliding friction of PDMS that exhibits a maximum at a similar velocity region. The third term of Eq. (14a) represents a simple viscous friction with a Langevin relaxation time $\tau_L$, which is obtained from the drift velocity ($\bar{\gamma}\tau_L$) in the limit of high $K$.

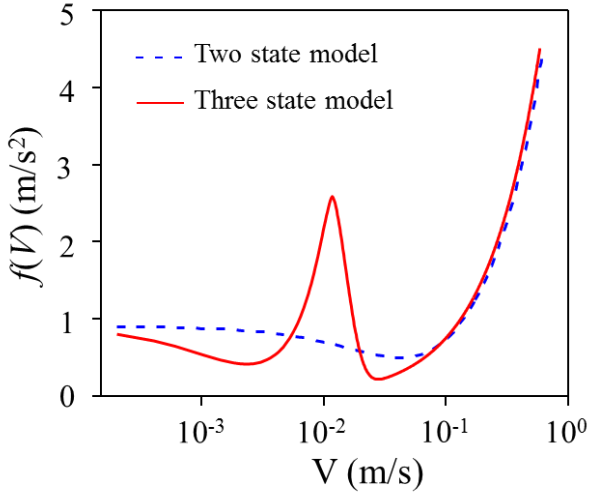

FIG. 11. (Color online) Approximate models of rolling friction versus velocity.

The parameters $V_o$, $A$ and $\sigma_V$ were obtained by a numerical fitting procedure, i.e matching the drift velocities at few values of $K$. Eq. (15) is similar to Eq. (1) with the difference that the acceleration term $dV/dt$ is multiplied by 7/5 which appears as a pre-factor to the acceleration when the equation of motion is derived from the balance of rotational torque and the derivative of angular momentum. $f(|V|)$ is the generalized friction force per unit mass, which is multiplied by $tanh(\alpha V)$ as this hyperbolic function with a high value of $\alpha$ ($\sim 10000$ s/m) is a good replacement for the signum function. Here, $\bar{\gamma}$ is 0.04 mN corresponding to the angle of inclination of $1^\circ$. Numerical solution of Eq. (15) was carried out using a generalized integration method of Gillepsie [36], as outlined before.

Before describing the friction model of Eq. (14), we first study a model in which a dry friction decreases exponentially with the velocity, i.e. $\Delta = \Delta_o \exp(-V/V_o)$, in conjunction with a viscous friction. Although such a model can reproduce the experimental drift velocity at different values of $K$ fairly well

Rolling motion of a sphere(Fig. 6), the spatial displacement statistics (Fig. 12) simulated with this model, however, are visibly sharp, with the sharpness persisting for a longer τ than that is observed experimentally.

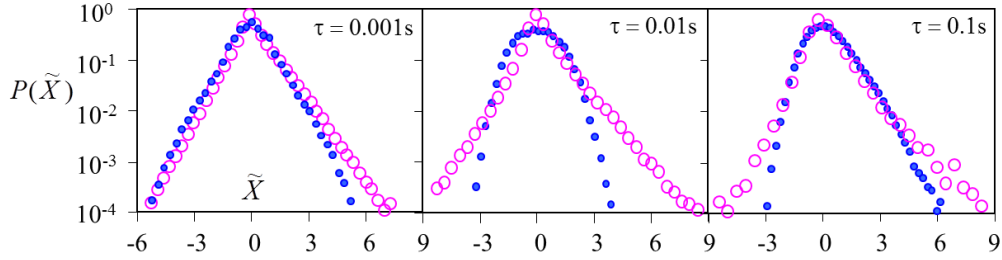

FIG. 12. (Color online) The pdf of the displacement fluctuation at different values of τ as obtained from experiments (filled blue circle) and from simulations (open pink circle) using the two state model of friction, in which the friction is described as $f(V) = \Delta_o \exp(-V/V_o) + V/\tau_L$. The values of $\Delta_o$, $V_o$ and $\tau_L$ are set as 0.9m/s², 0.028m/s and 0.13s respectively.

The simulated pdfs of displacement fluctuation obtained using the three state model of friction (Eq. (14)) are shown in Fig. 13. The essential features of the pdfs are, in general, consistent with the experimental observations. However, there are discrepancies. For example, the simulated pdf at $0.001s$ for the low power noise is not as fat tailed as that of the experiment. The exponent of the stretched Gaussian pdf is about 1 near the peak region, but it progressively increases to 1.7 near the tail region in comparison to overall exponent of ~ 1.2 obtained experimentally. Nevertheless, the sharpness of the pdf disappears faster with τ, and the transition to a smoother pdf occurs at a much shorter time scale than that predicted by the two state model of friction (Fig. 14). The simulated pdfs at the higher power of the noise are in better agreement with those obtained experimentally. Using the same friction model (eq. 14), we also estimated the drift velocities by integrating equation 15 at various values of $K$, the values of which are in satisfactory agreement with the experimental observations (Fig. 6).

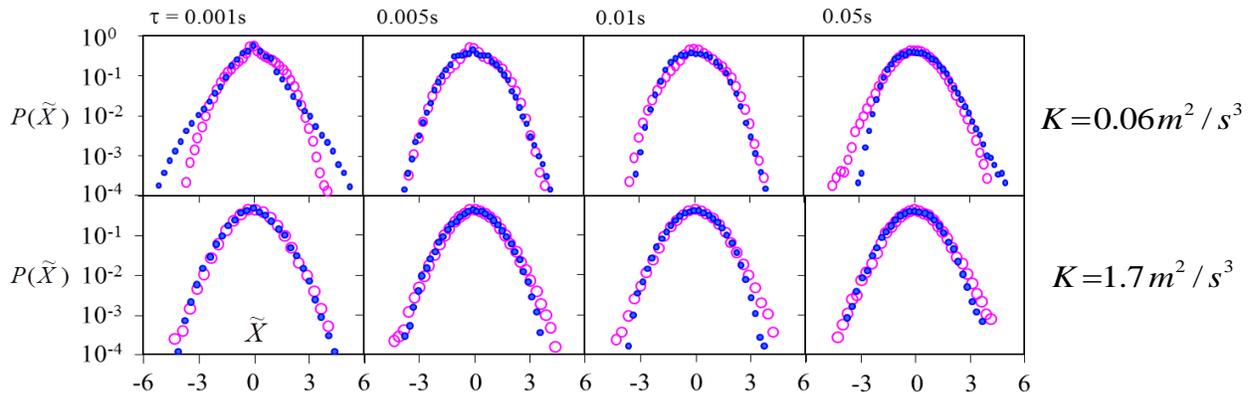

FIG. 13. (Color online) The pdf of the displacement fluctuation at different values of τ as obtained from experiments ( filled blue circle ) and from simulations ( open pink circle ) using the three state friction model (Eq. (14)).

## V. CONCLUDING REMARKS

This exploratory research revealed several interesting phenomenology of non-linear rolling friction. We summarize below the main points of the work and discuss what remains as open questions. The first point is about the Brownian like drift of the steel ball, which is linear in time in spite of the fact that the underlying frictional dynamics is non-linear. In the Einsteinian Brownian motion with a linear kinematic friction, it is an established fact that the object drifts linearly with time that being independent of the strength of the noise. However, in the current case of a non-Einsteinian Brownian motion, the drift velocity depends strongly on the strength of the noise. The rolling motion is controlled by a Coulomb like friction at low $K$, but by a viscous like friction at high $K$ suggesting, furthermore, a possible fluidization of the interface with noise. The pattern of the displacement pdfs suggests that a higher order non-linearity operates at an intermediate velocity, while a linear friction operates at even a higher velocity with the non-linearity weakening with $K$.

An evidence of a complex friction law comes from the observation of the drifted motion and the flow reversal of the ball when it is subjected to an asymmetric vibration. At a much higher amplitude of the vibration, the dynamics becomes far more complex, understanding that is the subject of our future study. The cusp singularity is, indeed, composed of multiple frequencies some of which could excite transverse vibrations of the fibrils. In future, this point should be examined by submitting the ball to accelerations of various frequencies and using fibrils of various physical characteristics. While the flow reversal indicates a descending and then an ascending branch of friction with velocity, it is not implausible that friction could depend on the magnitude of acceleration as well. This may not only manifest kinetically, there is already evidence that even its velocity independent component depends on the applied acceleration (or force). The data of Fig. 2 can indeed be interpreted as if the dry friction ($\Delta$) increases with $\bar{\gamma}$ non-linearly.

What is pertinent, however, to the point regarding the state dependent friction is that the drift velocity due to asymmetric vibration decreases significantly with the strength of the noise, which is contrary to what happens with a fixed bias. Taken together, the above evidences suggest that friction depends not only on the state ($K$) and the rate ($V$), but, possibly, could depend on the acceleration ($\dot{V}$) as well. In terms of developing a microscopic model of friction, we need to consider several factors. The first one being the rates at which interfacial bonds are formed and broken. It is also important to consider the roles of certain characteristic time scales of the fibrils, one of which comes from the ratio of the fibrillar spacing [11, 47] to the rolling speed and the other relates to the resonant frequency (~100 kHz) of the fibrils. The advantage of the current model system is that these parameters can be rigorously studied by careful design of the fibrillar geometry with which to develop a state and rate [48-50] dependent model of friction at any given value of $K$. Although a $K$ and $V$ dependent toy model of friction reproduces the essential features of the drift velocity and the evolution of the displacement pdf, some of the disagreements of the displacement statistics of the simulation and experiment clearly show that our understanding of the rolling friction dynamics is incomplete. A direct measurement of the rolling friction [10] spanning several decades of velocity and acceleration is very much needed in order to make further progress in this research. The value of the current work is that it could guide the designs of such experiments and set the stage for studying friction using the methods of statistical mechanics. If methods are developed to measure the statistics of the velocity fluctuations, then these data, in conjunction with the the displacement statistics, could be used for analyzing frictional dynamics more directly than that can be achieved with displacement statistics.

In the Langevin model, we tacitly assumed that the friction term has no memory. However, with the simultaneous presence of the elastic (due to fibrils) and viscous response of the system, friction may be viscoelastic. The elastic response of the fibrils, along with the non-linear friction dynamics may also exhibit spatio-temporal oscillations [8, 51, 52] in the rolling motion.

In addition to elucidating the frictional dynamics in detail, it will also be important to understand the nature of friction in the limit of vanishing $K$, which has not been clearly understood thus far. As pointed out above, the dry friction at $K=0$ depends on the applied acceleration. Furthermore, a low level of noise in conjunction with an asymmetric vibration increases the drift velocity initially, before it decreases with further increase of noise. Careful experiments and analysis of data utilizing systematic design of the patterned surfaces would be required in order to unfold the mysteries of Coulombic friction near zero noise limit.

We close this paper by pointing out that a stretched Gaussian displacement fluctuation must arise from a kinetic potential of the form $\mu(V) \sim |V|^m$, provided that the noise is Gaussian. This power law behavior of the kinetic potential leads to some anomalous work fluctuation relations [53, 54], which will be discussed in a separate publication.

## ACKNOWLEDGEMENTS

Discussions with M. O. Robbins and N. Ramakrishnan are gratefully acknowledged.

## References

[1] P. S. Goohpattader and M. K. Chaudhury, J. Chem. Phys. **133,** 024702 (2010).

[2] P. S. Goohpattader, S. Mettu and M. K. Chaudhury, Langmuir **25,** 9969 (2009).

[3] S. Mettu and M. K. Chaudhury, Langmuir **26,** 8131 (2010).

[4] B. Wang, S. M. Anthony, S. C. Bae and S. Granick, Proc. Natl. Acad. Sci. U. S. A. **106,** 15160 (2009).

[5] N. J. Glassmaker, A. Jagota, C. Y. Hui, W. L. Noderer and M. K. Chaudhury, Proc. Natl. Acad. Sci.U.S.A. **104,** 10786 (2007).

[6] K. A. Dahmen, Y. Ben-Zion and J. T. Uhl, Phys. Rev. Lett. **102,** 175501(2009).

[7] D. S. Fisher, Physics Reports **301,** 113 (1998).

[8] D. Maugis and M. Barquins, in *Adhesion 12*, edited by K. W. Allen (London, Elsevier, 1988), p. 205.

[9] M. Barquins and A. D. Roberts, J. Phys. D: Appl. Phys. **19,** 547 (1986).

[10] B. N. J. Persson, Euro. Phys. J. E **33,** 327 (2010).

[11] K. A. Grosch, Proc. R. Soc. London, Ser. A **274,** 21(1963).

[12] K. Vorvolakos and M. K. Chaudhury, Langmuir **19,** 6778 (2003).

[13] A. I. Leonov, Wear **141,** 137 (1990).

[14] A. E. Filippov, J. Klafter and M. Urbakh, Phys. Rev. Lett. **92,** 135503-1(2004).

[15] B. N. J. Persson, *Sliding Friction: Physical Principles and Applications* (Berlin: Springer, 2000).

[16] T. K. Caughey and J. K. Dienes, J. Appl. Phys. **32,** 2476 (1961).

[17] T. K. Caughey, J. Acoust. Soc. Am. **35,** 1706 (1963).

[18] J. M. Johnsen and A. Naess, in *Proc. EURODYN'93- Structural Dynamics*, edited by T. Moan et al (Balkema, Rotterdam, 1993) ISBN 90 5410336 1.

[19] G. Ahmadi, Int. J. Engng. Sci. **21,** 93 (1983).

[20] S. H. Crandall, S. S. Lee and J. H. Williams Jr., J. Appl. Mech. **41**, 1094 (1974)

[21] P. G. de Gennes, J. Stat. Phys. **119**, 953 (2005).

[22] A. Kawarada and H. Hayakawa, J. Phys. Soc. Japan **73,** 2037 (2004).

[23] A. Baule, E. G. D. Cohen and H. Touchette, Nonlinearity **24,** 351 (2011).

[24] H. Touchette, E. Van der Straeten and W. Just, J. Phys. A-Math. Theor. **43,** 445002 (2010).

[25] A. M. Menzel and N. Goldenfeld, Phys. Rev. E **84,** 011122 (2011).

[26] M. A. Moser and W. D. Iwan, J. Sound Vib. **159,** 223 (1992).

[27] M. K. Chaudhury and S. Mettu, Langmuir **24** 6128 (2008).

[28] J. A. Greenwood, K. L. Johnson, S-H. Choi and M, K. Chaudhury, J. Phys. D: Appl. Phys. **42,** 035301 (2009).

[29] S. Ratynskaia, G. Regnoli, K. Rypdal, B. Klumov and G. Morfill, Phys. Rev. E **80,** 046404 (2009).

[30] B. Lindner, New J. Phys. **9**, 136 (2007). Previously, Lindner suggested a similar scaling relation for diffusivity.

[31] T. Baumberger, L. Bureau, M. Busson, E. Falcon, and B. Perrin, Rev. Sci. Instrum. 69, 2416 (1998).

[32] I. Sánchez, F. Raynaud, J. Lanuza, B. Andreotti, E. Clement and I. S. Aranson, Phys. Rev. E **76,** 060301 (2007).

[33] M. Kardar, Phys. Rev. Lett. **55,** 2235 (1985).

[34] M. H. Muser, Proc. Natl. Acad. Sci. U.S.A. **107,** 1257 (2010).

[35] L. Z. Prandtl, Angew Math. Mech. **8,** 85 (1928).

[36] D. T. Gillespie, Am. J. Phys. **64,** 225 (1996).

[37] S. Daniel, M. K. Chaudhury and P-G. de Gennes, Langmuir **21,** 4240 (2005).

[38] D. Fleishman, Y. Asscher and M. Urbakh, J. Phys.: Condens. Matter **19,** 096004 (2007). These authors suggested the possibility of flow reversal with a non-linear sliding friction coupled to asymmetric vibration.

[39] P. Brunet, J. Eggers and R. D. Deegan, Phys. Rev. Lett. **99,** 144501 (2007).

[40] M. Eglin, M. A. Eriksson and R. W. Carpick, J. Appl. Phys. **88,** 091913 (2006).

[41] A. Buguin, F. Brochard and P.-G. de Gennes, Eur. Phys. J.E **19,** 31 (2006).

[42] S. Mettu and M. K. Chaudhury, Langmuir ( http://pubs.acs.org/doi/abs/10.1021/la201597c ).

[43] O. Zik, J. Stavans and Y. Rabin, Europhys. Lett. **17,** 315 (1992).

[44] G. D'Anna, P. Mayor, A. Barrat, V. Loreto and F. Nori, Nature **424,** 909 (2003).

[45] Here, "state" means whether the system is in a solid-like or a liquid-like state. At any given level of noise, friction depends on various variables, leading to the well-known "state and rate" law of friction.

[46] S. Andersson, A. Soderberg, S. Bjorklund, Tribol. Int. **40**, 580 (2007).

[47] E. Wandersman, R. Candelier, G. Debregeas, and A. Prevost, arxiv.org/PS_cache/arxiv/pdf/1107/1107.2578v1.pdf

[48] J. R. Rice and A. L. Ruina, J. Appl. Mech. **50,** 343 (1983).

[49] J. H. Dieterich, J. Geophys. Res. **84,** 2161 (1979).

[50] J. R. Rice, N. Lapustaa and K. Ranjith, J.Mech. Phys. Sol. **49,** 1965(2001).

[51] M. Morishita, M. Kobayashi, T. Yamaguchi and M. Doi, J. Phys.: Condens. Matter **22,** 365104 (2010).

[52] R. De, A. Maybhate and G. Ananthakrishna, Phys. Rev. E **70,** 046223 (2004).

[53] H. Touchette and E. G. D. Cohen, Phys. Rev. E **80**, 011114 (2009).

[54] A. V. Chechkin and R. Klages, J. Stat. Mech.: Theory Exp. **2009**, L03002 (2009).